\documentclass[fleqn,10pt]{wlscirep}
\usepackage{chngcntr}
\usepackage[utf8]{inputenc}
\usepackage[T1]{fontenc}
\usepackage{makecell}

\title{Probing depressive symptoms and the desire to leave academia among scientists in large, international collaborations in STEM}

\author[1,2,*]{Kamiel Janssens }
\author[3,4]{Michiko Ueda}
\affil[1]{Universiteit Antwerpen, Prinsstraat 13, 2000 Antwerpen, Belgium}
\affil[2]{Universit\'e C\^ote d’Azur, Observatoire de la C\^ote d’Azur, CNRS, ARTEMIS, 06304 Nice, France}
\affil[3]{Department of Public Administration and International Affairs, Syracuse University, Syracuse, New York, United States}
\affil[4]{Center for Policy Research, Maxwell School of Citizenship and Public Affairs, Syracuse University, Syracuse, New York, United States}

\affil[*]{kamiel.janssens@uantwerpen.be}

\begin{abstract}

Large-scale international scientific collaborations are increasingly common in the field of STEM (Science, Technology, Engineering, and Mathematics). However, little is known about the well-being of the members participating in these `big science' collaborations, which can present unique challenges due to the 
scale of their work.  We conducted a survey among members of three large, international collaborations in the field of gravitational-wave astrophysics in the summer of 2021. Our objective was to investigate how career stage, job insecurity and minority status are associated with reported levels of depressive symptoms as well as the desire to leave academia. We found that early-career scientists and certain minoritized groups reported significantly higher levels of depressive symptoms compared to senior members or those who do not consider themselves as a member of minoritized groups. Furthermore, relatively young members, staff scientists/engineers, and those experiencing high levels of job insecurity and lack of recognition were more likely to frequently consider leaving academia. Our findings suggest that improving recognition for personal contributions to collaborative work and providing clearer job perspectives could be two key factors in enhancing the well-being of young scientists and reducing the potential outflow from academia.

\end{abstract}
\begin{document}

\flushbottom
\maketitle
\thispagestyle{empty}

\section{Introduction}
\label{sec:introduction}

Many present-day research projects in STEM (Science, Technology, Engineering, and Mathematics) require large investments both concerning funding as well as person power. Therefore it is often no longer feasible for a single institution to conduct such research. Instead a large number of institutions join effort in a scientific collaboration to which they contribute with financial resources and/or person power \cite{Borner:2021}.
In addition to their large membership size, members of these collaborations are typically scattered across multiple continents and time zones. This creates unique working conditions in terms of working hours, work-related pressure, and recognition for individual efforts. Some of these factors could potentially contribute to additional stress on mental health. For instance, in large collaborations, published papers are often attributed to the entire collaboration, involving hundreds or even thousands of authors, without specifically identifying individual contributions. This may impact members' sense of recognition for their individual efforts within the collaboration. Additionally, due to the global nature of these collaborations, online meetings often fall outside regular office hours. Furthermore, these international collaborations can encompass a culturally diverse group of members, presenting challenges in creating a safe, culturally sensitive, and inclusive work environment for all \cite{Lucatello:2017}. On the other hand, these collaborations can also foster an active network where individuals closely work with a large number of collaborators. For instance, in the scientific collaborations examined in this study, social events for early-career researchers are actively organized during large internal meetings held twice a year. These networks of collaborators can foster a sense of solidarity among scientists in these collaborations, which may not be available to scientists who are not part of such collaborative projects.

To the best of our knowledge, no study has examined the well-being and desire to leave academia among members of large, international scientific collaborations. However, previous studies have indicated that certain groups within the scientific community are prone to experiencing mental health challenges. For instance, earlier research has shown that PhD students often face poor mental health \cite{Williams2014,ESSADEK2020392, Evans:2018}. Evans et al. \cite{Evans:2018} reported high levels of depressive and anxiety symptoms among graduate students (both doctoral and master's students) in STEM and other academic domains. The unfavorable work-life balance and compromised mental well-being can result in young researchers leaving the academic environment shortly after completing their PhD, as demonstrated by various previous studies \cite{Metcalfe:2018,Cornell:2020,Hancock:2020}. The significant outflow of young researchers could potentially impact the quality of future research.

Another study examined how the research culture affects the mental health of STEM researchers and highlighted the detrimental effects of job insecurity and the competitive academic environment on their psychological well-being \cite{Limas:2022}. Moss et al. \cite{Moss:2022} investigated the relationship between mental health literacy, help-seeking behavior, and levels of psychological distress among postgraduate researchers at two UK universities. They found that lower levels of well-being were associated with higher levels of distress and decreased help-seeking \cite{Moss:2022}.

Furthermore, there is evidence suggesting that the psychological well-being of researchers is worse than that of the general population or other highly educated populations. Researchers commonly experience higher levels of burnout compared to the general population, aligning more closely with `high-risk' employees such as healthcare workers \cite{Guthrie:2017}. Moreover, the study found that a greater proportion of researchers self-report mental health problems compared to the general population. A study conducted in the United Kingdom found that doctoral researchers reported significantly higher levels of depressive and anxiety symptoms compared to similarly highly educated individuals in non-academic positions, even after controlling for various confounding factors \cite{Hazell:2021}. Additionally, Leveque et al. \cite{LEVECQUE2017868} demonstrated that PhD students in Belgium exhibit a higher prevalence of mental health problems compared to comparable groups of highly educated individuals.

Furthermore, the COVID-19 pandemic disrupted the daily lives of many individuals and had a significant impact on their mental well-being \cite{ESSADEK2020392,FANCOURT2021141,Shevlin2020,Palgi2020,Ramiz2021}. The pandemic may have exacerbated the existing inequalities observed in academia. Emerging evidence indicates that the pandemic had a negative impact on the productivity of early-career and female researchers across various disciplines, including STEM \cite{Bohm_Liu_2023, Squazzoni_2021, Heo_2022}.

In this study, our objective is to investigate the reported levels of depressive symptoms and the desire to leave academia among researchers involved in large, international collaborations in the STEM field. To achieve this, we conducted an online survey during the summer of 2021, which can be considered as the post-COVID-19 pandemic period in many member countries, including the United States. However, for certain countries in the collaboration, notably Japan, COVID-related restrictions were not fully relaxed until May 2023, making them still in the midst of the ongoing pandemic during the survey. Specifically, our survey targeted scientists in the field of gravitational-wave astrophysics who are members of the LIGO (Laser Interferometer Gravitational-Wave Observatory), Virgo, and KAGRA (Kamioka Gravitational Wave Detector) collaborations. These collaborations operate gravitational-wave detectors located in the United States, Europe, and Japan, respectively. 
The respective sizes of the different collaborations are: LIGO $\sim$ 1450 members, Virgo $\sim$ 470 members and KAGRA $\sim$ 450 members. Although the collaborations represent three separate entities, each with their own gravitational-wave detector(s), they have come together through a memorandum of agreement due to their shared scientific goals and challenges, namely, the detection of gravitational waves. The exploration of the scientific data is therefore conducted jointly.

By conducting this first well-being survey focused on large and international collaborations in the STEM field, our aim is to explore the prevalence of depressive symptoms and the intention to leave academia or the collaboration among its members. In particular, we investigate whether early-career scientists, who often face uncertain job prospects, experience different levels of depressive symptoms and exhibit a greater inclination to leave academia compared to senior, tenured faculty members who enjoy relatively secure positions and possibly greater work flexibility. Additionally, this project examines whether members who identify as belonging to minoritized groups, either within the collaboration or the country of their residence, experience distinct levels of psychological well-being and a differing willingness to leave academia or collaborative efforts compared to their peers who do not identify as belonging to minoritized groups. Lastly, we aim to understand the relationship between the high levels of job insecurity that might be experienced by early-career scientists and their mental health, as well as their desire to leave the academic field.

\section{Methods}
\label{sec:methods}

\subsection{Recruitment and participants}

Data collection for the survey started on 23rd August 2021 and concluded on 30th September 2021. An initial email to announce the survey was sent at the start of data collection to all members of the collaborations through the general email lists of the LIGO, Virgo, and KAGRA collaborations, to which all active members are expected to be subscribed. Additionally, the survey was announced during two collaboration meetings: a KAGRA internal meeting (27 - 29 August 2021) and a joint LIGO-Virgo-KAGRA collaboration meeting (6 - 10 September 2021). Finally, a reminder email was distributed to the same mailing lists several days before the end of the data collection period.

The survey was conducted online and maintained complete anonymity. This is the first well-being survey conducted by  these collaborations. Participation in the survey was voluntary, and no monetary compensation was provided to the participants. To ensure confidentiality and create a safe environment for freely expressing their opinions, it was clearly stated at the beginning of the survey that only the research team would have access to the data. Furthermore, it was explicitly mentioned that participation in the survey would not have any effect on participants' current or future careers. All participants included in the analysis explicitly provided their informed consent. However, five participants were excluded from the subsequent analysis as they had skipped one or more crucial questions necessary for the analysis. An additional 15 participants were excluded because they were not actively engaged in research (e.g. administrative staff, retired professors). 
As a result, our final sample consisted of 397 participants. Ethical approval for the study was obtained from the ethical review committee of the University of Antwerp (reference number: SHW\_21\_75).

The survey was initially created in English. To increase participation rates from members in Asia, we also made the survey available in Japanese, Chinese (both traditional and simplified characters), and Korean. The distribution of respondents across the languages is as follows: English: 339 (82.3\% of all participants), Japanese: 65 (15.8\%), Chinese (Simplified and Traditional combined): 5 (1.2\%), and Korean: 3 (0.7\%).

Participants indicated their primary affiliation, and the sample sizes for each collaboration were as follows: LIGO: 202 (49.0\% of all participants), Virgo: 109 (26.5\%), and KAGRA: 89 (21.6\%). Some participants' primary affiliations were not one of the three targeted collaborations in this survey, and they were mainly affiliated with different collaborations in related research fields. This group accounted for the remaining 12 (2.9\%) participants.

The response rate for each collaboration can be estimated using two different methods. One approach would be to consider the number of individuals subscribed to the mailing lists through which the survey was distributed. However, it should be noted that these lists may include individuals with varying levels of activity within the collaboration, thus providing a lower limit for the participation rate. Another method involves focusing solely on individuals listed as authors within the collaboration, as this represents a more restricted and active group of members. The lower (upper) limits of the participation rates are as follows: LIGO: 14.2\% (21.3\%), Virgo: 23.2\% (24.4\%), and KAGRA: 20.2\% (41.7\%), respectively.

\subsection{Measures}

\paragraph{Depressive symptoms}\hspace{1cm} 

Depressive symptoms were measured by using the Patient Health Questionaire (PHQ-9), which is a 9-item set of questions used for diagnosing depressive symptoms \cite{PHQ-9}. Participants respond to the questions using a four-point Likert scale ranging from 0 (not at all) to 3 (nearly every day), focusing on their experiences of the last two weeks. The total score is ranged between 0 and 27, where high scores indicate higher levels of depressive symptoms. An often used criteria for people experiencing depressive symptoms is a cut-off score of 10 \cite{PHQ-9}. The internal consistency between the different questions of PHQ-9 was good with a Cronbach's alpha value of 0.89 (95\% confidence interval (CI): [0.87 ; 0.90]). 

\paragraph{Leaving academia/collaboration}\hspace{1cm} 

Participants were asked how often they considered to leave academia/the collaboration in the last 9 months. They could choose one of the following options: (1) Never, (2) Once or twice, (3) Several times per month, (4) Several times per week, (5) Every day, and (6) Prefer not to answer. We created an indicator variable that takes a value of 1 if they indicated that they considered leaving academia/the collaboration more than several times per week (i.e. those who chose (4) and (5)), which captures their desire to leave academia/the collaboration. 

\paragraph{Demographics} \hspace{1cm} 

Self reported age was recorded and divided in four categories. The age 50+ category will be used as the reference group to which different age categories are compared. As for gender identity, respondents could indicate one or multiple of the following categories: (1) male, (2) female, (3) cisgender, (4) transgender, (5) non-binary, (6) non-queer, (7) other, and (8) prefer not to answer. For the subsequent analysis, people indicating (fe)male or (fe)male + cisgender were categorized as `(Fe)male'. Other participants were grouped in the `Other' category. In the regression analysis, male respondents are used as the reference group, to which female and those whose gender identify is `other' will be compared. 

\paragraph{Career level}\hspace{1cm} 

Participants were asked to indicate their career level, for which they could choose from: (1) Bachelor (BA)/Master (MA) student, (2) PhD student, (3) post-doc, (4) staff scientist/engineer, (5) tenure track professor, (6) tenured professor, (7) other. As noted above, those who selected `other' were excluded from the analysis. Tenured professors are set as the reference group in the regression analysis. 

\paragraph{Working location}\hspace{1cm} 

Participants were asked whether they primarily work on-site at the detector or in an office. Given the global measures to fight the spread of the COVID-19 virus, people who were not working at the detector sites are considered to be more likely to have been working form home due to the less crucial character of their `on-site' presence. In the regression analysis, those who primarily work on-site are set as the reference group. 

\paragraph{Uneven workload}\hspace{1cm} 

We asked `How often is the workload unevenly distributed so that it piles up?'. Participants could answer: Never (1), Rarely (2), Sometimes (3), Often (4) and Always (5). For the analysis, the first two options are coded as the `Low' level of uneven workload, the third item (`Sometimes') as `Medium' and the last two as `High'. Those who reported a low level of uneven workload are set as the reference group in the regression analysis.

\paragraph{Work-life interference}\hspace{1cm} 

We asked two questions pertaining work-life balance/interference. `How often do you feel that the demands of your work interfere with your private or family life?' and `How often do you feel the irregular meeting times interfere with your private or family life?' The latter question was asked as meeting times often can fall outside the regular office hours due to the global character of the joined research performed in collaboration between LIGO, Virgo and KAGRA.
Participants could answer: Never (1), Rarely (2), Sometimes (3), Often (4) and Always (5). Both question were summed to get one overall parameter which serves as an indicator for work-life interference and can take values between 2 (low levels of interference) and 10 (high levels of interference). The two questions had a Cronbach's alpha of 0.76 (95\% CI: [0.71 ; 0.83]), indicating a decent internal consistency. 
For the analysis, a combined score of 2-4  was categorized as the `Low' level of work-life interference, a score of 5-7 as `Medium' and a score of 8-10 as `High'. Those who reported a low level of work-life interference are set as the reference group in the regression analysis. 

\paragraph{Job influence}\hspace{1cm} 

Participants were asked whether they have a large degree of influence on the decisions concerning their work and could answer with one of the following options: (1) Never, (2) Rarely, (3) Sometimes, (4) Often, and (5) Always. Items (1) and (2) were grouped as `Low', (3) as `Medium' and (4) and (5) as `High' levels of job influence in the analysis. Those who reported `High' level of influence in their work are set as the reference group in the regression analysis. 

\paragraph{Recognition for work}\hspace{1cm} 

Participants were asked whether they think their work as being part of the collaboration is properly recognised. They could respond with one of the following options: (1) Yes, both inside and outside the collaboration, (2) Only inside the collaboration, (3) Only outside the collaboration, (4) Neither inside nor outside the collaboration. We coded (1)-(3) as `Yes' and (4) as `No' with respect to recognition, where the former is used as reference group in the subsequent analysis. 

\paragraph{Meaningfulness of work}\hspace{1cm} 

Participants were asked how meaningful they find their work and were able to choose between one of the following options: (1) To a very large extent, (2) To a large extent, (3) Somewhat, (4) To a small extent and (5) To a very small extent. For the analysis, the first two options are coded as the `High' level of meaningfulness, the third item (`Somewhat') as `Medium' and the last two as `Low'. Those who reported a high level of meaningfulness are set as the reference group in the regression analysis. 

\paragraph{Job insecurity}\hspace{1cm} 

Participants were asked whether they felt insecure about the future of their job and could answer with one of the following options: (1) Never, (2) Rarely, (3) Sometimes, (4) Often and (5) Always. Options (1) and (2) were grouped in `Low', (3) in `Medium' and (4) and (5) in `High' levels of job insecurity for the subsequent analysis. The `Low' group will be used as reference category. 

\paragraph{Minoritized groups}\hspace{1cm} 

Participants were asked whether they considered themselves as part of a minoritized group and could answer (1) Yes, in my country of residence, as well as the collaboration, (2) Yes, in the collaboration, (3) No and (4) Prefer not to answer. People who indicated to consider themselves to be part of a minoritized group were subsequently asked several additional questions. Non-minoritized participants are the reference group in the regression analysis.

\section{Results}
\label{sec:results}

The prevalence of depressive symptoms and the proportion of individuals seriously contemplating leaving academia or the collaboration are presented in Table \ref{tab:Prevalence}, alongside the corresponding number of respondents in each category. Among the participants in this survey, the prevalence of depressive symptoms, as indicated by a PHQ-9 cut-off score of 10 or higher (indicating moderate to severe symptoms), was found to be 22.92\% (95\% CI: 18.88\%-27.38\%). The mean PHQ-9 score was 6.69 with a standard deviation of 5.86. Furthermore, 18.90\% (95\% CI: 15.16\%-23.09\%) of the participants reported considering leaving academia or the collaboration several times a week or more.

The results presented in Table \ref{tab:Prevalence} indicate notable variations in the prevalence of depressive symptoms based on participants' career stages, with early-career scientists exhibiting higher levels of depressive symptoms compared to their more senior counterparts. To explicitly examine this difference, we further analyzed the prevalence of depressive symptoms separately for early-career scientists (BA/MA students, PhD students, and post-docs) and more senior scientists (staff scientists/engineers, tenure-track professors, and tenured professors). The prevalence among early-career scientists was found to be 32.8\% (95\% CI: 26.45\%-39.75\%), while among more senior scientists, it was 12.4\%(95\% CI: 8.13\%-17.94\%). This difference between the two groups was statistically significant, with $t$=5.02 (p-value=8.15e-7), as determined by a two-sided Welch's t-test ($\alpha=0.05$).

Regarding the desire to leave academia or the collaboration, the findings in Table \ref{tab:Prevalence} indicate that individuals who report a lack of recognition, influence, and meaningfulness in their work are more likely to seriously consider leaving. Furthermore, the results also suggest that those who face challenges in achieving a satisfactory work-life balance tend to contemplate exploring alternative career paths.

As previously mentioned, participants had the option to indicate whether they considered leaving only the collaboration, only academia, or both. Among the 62.7\% of participants ($N$=249) who stated that they had considered leaving the collaboration and/or academia at least once, 47.8\% expressed thoughts of leaving both academia and the collaboration, 27.7\% considered leaving only academia, 20.1\% contemplated leaving only the collaboration, and 4.4\% preferred not to specify whether they wanted to leave academia or the collaboration.

\begin{figure}
    \centering
    \includegraphics[width=\linewidth]{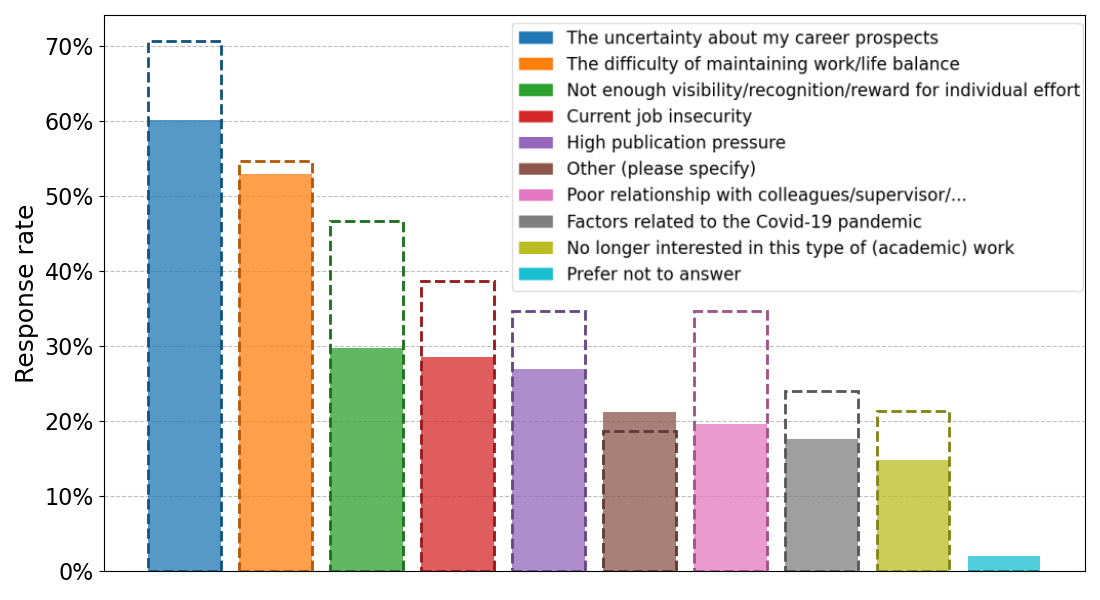}
    \caption{Histogram indicating the different reasons why people consider leaving the collaboration/academia (filled bins). 249 people considered at least once to leave the collaboration/academia. On average each respondent gave about 2.7 reasons, leading to a total of 674 answers to the reasons for leaving shown here. The shown percentages are with respect to the number of people who responded to this question.
    The dashed bin edges represent the response rates of the 75 participants who consider leaving at least several times a week. These participants gave on average 3.4 reasons for considering to leave academia/the collaboration.}
    \label{fig:ReasonsLeaving}
\end{figure}

Participants were also asked to indicate the reason(s) why they wanted to leave academia or the collaboration, with the option to select multiple reasons. The distribution of the 674 indicated reasons is illustrated in Fig. \ref{fig:ReasonsLeaving} (filled bins), where the percentages are calculated based on the number of participants considering leaving academia or the collaboration. As participants could choose more than one reason, the sum of the percentages exceeds 100\%. As shown in Fig. \ref{fig:ReasonsLeaving}, the two most common reasons selected were `\textit{The uncertainty of my career prospects}' and `\textit{The difficulty of maintaining work-life balance}', which were cited by over half of the participants who contemplated leaving academia or the collaboration. In Fig. \ref{fig:ReasonsLeaving} we also show the response for a subset of participants who consider regularly (at least several times a week) to leave academia/the collaboration (dashed bin edges).

While the results presented in Table \ref{tab:Prevalence} indicate that certain groups are more likely to experience depressive symptoms and contemplate leaving their current position than others, it is important to consider other potential factors that may explain these observed patterns. To account for these factors, we conducted a multivariate linear regression analysis with the PHQ-9 score as the dependent variable to examine the correlates of depressive symptoms. Additionally, we estimated a multivariate logit model with a binary indicator capturing the participants' desire to leave the collaboration/academia as the dependent variable. Both models included the same set of demographic, career-related, and work-related variables as independent variables. To account for potential heteroskedasticity, we used robust standard errors in all estimations \cite{MACKINNON1985305}.

Table \ref{tab:PHQ9} reports the regression results when the score for PHQ-9 was used as the dependent variable. The estimated coefficients are shown with the 95\% CI and the associated p-values. We consider predictor variables to be statistically significant if their p-value is $\leq$ 0.05, in which case they are shown in bold in Table \ref{tab:PHQ9} and \ref{tab:Leaving}. In Appendix \ref{sec:Appendix_logitPHQ9}, we present the estimation results for a logistic regression model in which the dichotomous dependent variable that captures the existence of depressive symptoms was used as the dependent variable. As in Table  \ref{tab:Prevalence}, we used a cutoff score of $\geq$10 for the PHQ-9 as an indication of moderate to severe depressive symptoms. 

As reported in Table \ref{tab:PHQ9}, we find that having high levels of job insecurity and being an early career scientist are strongly associated with higher levels of depressive symptoms. On average, the PHQ-9 score of people who feel insecure about their future job prospects tend to be higher by 3.71 (95\% CI: 2.22-5.19) compared to the reference group, even after controlling for other confounding factors. As reported above, the mean PHQ-9 score for all samples was 6.69 and the range of the score is between 0 and 27. 
In addition, the average PHQ-9 score for BA/MA students (3.67; 95\% CI: 0.68-6.65) as well as for PhD students (2.38; 95\% CI: 0.26-4.50) and staff scientists/engineers (1.52; 95\% CI: 0.06-2.98) are higher than the reference group, and these differences are all statistically significant. Furthermore, people who reported highly uneven workload tend to a have higher scores (3.52, 95\% CI: 1.97-5.07) as well as those who reported high levels of work-life interference (2.05, 95\% CI: 0.14-3.97), compared to the reference group who did not have any issues related to workload or work-life balance. Moreover, those that feel they lack recognition for their work tend to report higher PHQ-9 score (2.40; 95\% CI: 0.52-4.29), indicating stronger depressive symptoms.
Finally, participants who indicate to be part of a minoritized group both within their country of residence and the collaboration tend to record higher PHQ-9 scores (1.93, 95\% CI: 0.39-3.47). The breakdown of the minoritized status is shown in Appendix \ref{sec:Appendix_minority}. At the same time, we did not find any evidence that female members and those who identify themselves as `other' gender exhibit higher depressive symptoms compared to male counterparts. 
Further investigation revealed that post-docs and to a lesser extent PhD students and staff scientists, engineers struggle most for a lack of recognition, as well as participants from a minoritized group (see Appendix \ref{sec:Appendix_recognition}).

As for their desire to leave academia/the collaboration, Table \ref{tab:Leaving} reports the odds ratios (ORs) when we estimated a logit model with an indicator variable for the participants' desire to leave the collaboration/academia as the dependent variable.

Relatively young scientists (age 18-29) and staff scientists/engineers are the most likely to regularly consider leaving academia/the collaboration. Participants who are less than 30 years old are 21.57 times (95\% CI: 2.39-194.59) more likely to think weekly/daily about leaving academia/the collaboration, compared to their more senior counterparts. Staff scientists/engineers are 4.02 times (95\% CI: 1.22-13.24) more likely to think regularly about leaving the collaboration compared to tenured professors. 
In addition, scientists who feel high levels of insecurity about their future job prospects (OR=3.87; 95\% CI: 1.62-9.26) and those do not feel properly recognized for their work (OR=2.90; 95\% CI: 1.33-6.31) are more likely to think about leaving academia/the collaboration. Finally, participants who have little influence on work-related decisions are 2.64 times (95\% CI: 1.11-6.26) more likely to consider leaving academia compared to their peers who can exercise a larger influence in their work.

\begin{table*}[h]
    \centering
    \begin{tabular}{|p{4,5 cm}p{2,5 cm}p{3.5 cm}p{4.5 cm}|}
        \hline
         & &  \multicolumn{2}{c|}{Prevalence}\\
         \hline
         & N (\%) & Depressed (PHQ-9$\geq 10$) & Leaving ($>$several times a week)\\ \hline
          Total & 397 (100\%)& 22.92\% & 18.90\%   \\
          Age & & &   \\
         \leavevmode\phantom{aaa} 18-29 & 114 (28.7\%) & 35.09\%  &    28.07\%  \\
         \leavevmode\phantom{aaa} 30-39 & 109 (27.5\%) & 23.85\%  &  22.02\%     \\
         \leavevmode\phantom{aaa} 40-49 &59 (14.9\%) & 11.86\% &    10.17\% \\
         \leavevmode\phantom{aaa} 50+ & 55 (13.9\%) &  12.73\% &  1.82\% \\
         \leavevmode\phantom{aaa} Unknown & 60 (15.1\%) & 18.33\%  &    20.00\%   \\
         Gender & & &   \\
         \leavevmode\phantom{aaa} Male &   273 (68.8\%)& 21.25\% & 17.58\%   \\
         \leavevmode\phantom{aaa} Female &  101 (25.4\%) & 26.73\%   &  21.78\%    \\
         \leavevmode\phantom{aaa} Other &   23 (5.8\%)  & 26.09\%  &    21.74\%   \\
         Career & &  &  \\
         \leavevmode\phantom{aaa}  BA/MA student &  21 (5.3\%) &  42.86\%  &     23.81\%   \\
         \leavevmode\phantom{aaa} PhD &  96 (24.2\%) & 35.42\%  &    22.92\%   \\
         \leavevmode\phantom{aaa} Post-doc & 87 (21.9\%) & 27.59\%  &    31.03\%  \\
         \leavevmode\phantom{aaa} Staff scientist, engineer & 65 (16.4\%)  & 18.46\% & 18.46\%   \\
         \leavevmode\phantom{aaa} Tenure track professor &  28 (7.1\%) & 14.29\%  &    14.29\%   \\
         \leavevmode\phantom{aaa} Tenured professor & 100 (25.2\%)  & 8.00\%  &    5.00\%   \\ 
         Working location: onsite  & & &  \\
         \leavevmode\phantom{aaa} Yes & 65 (16.4\%) & 16.92\%   &    12.31\%  \\
         \leavevmode\phantom{aaa}  No &  332 (83.6\%)  &  24.10\%   &   20.18\%  \\
         Uneven workload  & &    &             \\    
         \leavevmode\phantom{aaa} Low & 60 (15.1\%)  & 10.00\%  & 16.67\%      \\
         \leavevmode\phantom{aaa} Medium & 133 (33.5\%) & 18.80\%  &    15.79\%    \\         
         \leavevmode\phantom{aaa}  High & 204 (51.4\%)  & 29.41\%  &    21.57\%   \\            
         Work-life interference  & &   &          \\  
         \leavevmode\phantom{aaa} Low & 82 (20.7\%)  & 14.63\%  & 14.63\%     \\
         \leavevmode\phantom{aaa} Medium & 208 (52.4\%) & 20.67\%  &    14.90\%   \\        
         \leavevmode\phantom{aaa}  High & 107 (27.0\%)  & 33.64\%  &    29.91\%    \\               
         Job influence   & & &  \\
         \leavevmode\phantom{aaa} Low & 53 (13.4\%)  & 37.74\% &  33.96\%      \\
         \leavevmode\phantom{aaa} Medium & 117 (29.5\%) & 31.62\%  &   25.64\%  \\
         \leavevmode\phantom{aaa} High &227 (57.2\%)  & 14.98\%  &    11.89\%   \\         
         Recognition   & & &   \\
         \leavevmode\phantom{aaa} Yes & 345 (86.9\%) & 18.84\%  &    14.20\%   \\         
         \leavevmode\phantom{aaa}  No &  52 (13.1\%) &  50.00\%  &     50.00\%   \\           
         Job meaningfulness  & & &  \\
         \leavevmode\phantom{aaa} Low & 35 (8.8\%) & 40.00\% &      40.00\%   \\
         \leavevmode\phantom{aaa} Medium & 113 (28.5\%) & 30.97\%  &      25.66\%  \\         
         \leavevmode\phantom{aaa} High & 249 (62.7\%) & 16.87\%  &       12.85\%  \\            
         Job insecurity  & &  &  \\
         \leavevmode\phantom{aaa} Low & 165 (41.6\%)  & 9.09\%  &  7.27\%       \\
         \leavevmode\phantom{aaa} Medium & 85 (21.4\%) & 16.47\% &      10.59\%  \\         
         \leavevmode\phantom{aaa}  High &  147 (37.0\%) &    42.18\%  &    36.73\% \\           
         Minority status  & & &  \\
         \leavevmode\phantom{aaa} No & 271 (68.3\%) & 22.51\%    &  15.13\%  \\
         \leavevmode\phantom{aaa} Country \& collaboration & 66 (16.6\%)  &    28.79\%   & 24.24\%  \\        
         \leavevmode\phantom{aaa} Collaboration & 40 (10.1\%)  & 20.00\%  &    30.00\%   \\ 
         \leavevmode\phantom{aaa} Unknown & 20 (5.0\%) &  15.00\% &    30.00\%    \\         
\hline
    \end{tabular}
    \caption{Prevalence of depressive symptoms as measured by \textit{PHQ-9} (Patient Health Questionnaire-9) and the desire to leave academia by sample characteristics.}
    \label{tab:Prevalence}
\end{table*}

\begin{table*}[h]
    \centering
    \begin{tabular}{|p{4,6 cm}p{2 cm}p{2.75 cm}p{2 cm}|}
        \hline
         & \multicolumn{3}{c|}{Multivariate regression}\\
         \hline
         & Coeff & 95\% CI & p \\
         \hline
         Age &  & &  \\
         \leavevmode\phantom{aaa} 18-29 &  0.40  &    [-1.80, 2.60] &   0.721  \\
         \leavevmode\phantom{aaa} 30-39 &  -0.02  &    [-1.59, 1.55] &   0.980  \\
         \leavevmode\phantom{aaa} 40-49 &   -0.60  &   [-2.35, 1.15]  &   0.502  \\
         \leavevmode\phantom{aaa} 50+ &  Ref.& Ref.& Ref. \\
         \leavevmode\phantom{aaa} Unknown &-0.00  &    [-1.80, 1.79]  &   0.997  \\
         Gender  & & &  \\
         \leavevmode\phantom{aaa} Male &   Ref.& Ref.& Ref. \\
         \leavevmode\phantom{aaa} Female &   -0.06  &    [-1.26, 1.14]   &  0.917  \\
         \leavevmode\phantom{aaa} Other &   1.21  &    [-1.55, 3.97]  &  0.390  \\
         Career  & & &  \\
         \leavevmode\phantom{aaa} \bf BA/MA student  & \bf 3.67  & \bf  [ 0.68, 6.65] & \bf  0.016  \\
         \leavevmode\phantom{aaa} \bf PhD  &\bf 2.38  & \bf   [ 0.26, 4.50] & \bf  0.028  \\
         \leavevmode\phantom{aaa} Post-doc  &   -0.06  &   [-1.86, 1.74] &   0.951  \\ 
         \leavevmode\phantom{aaa} \bf Staff scientist, engineer   &\bf 1.52 & \bf[ 0.06, 2.98]   &\bf   0.041  \\ 
         \leavevmode\phantom{aaa} Tenure track professor  &  0.09  &    [-2.10, 2.28]  &   0.937  \\ 
         \leavevmode\phantom{aaa} Tenured professor   & Ref. &    Ref.&    Ref. \\         
         Working location:  onsite   & & &  \\
         \leavevmode\phantom{aaa} Yes  & Ref. &    Ref.&    Ref. \\
         \leavevmode\phantom{aaa} \bf No & \bf 1.10  & \bf   [ 0.00, 2.20] &\bf   0.050  \\
         Uneven workload  &     &      &       \\    
         \leavevmode\phantom{aaa} Low   & Ref. & Ref.   &  Ref.  \\
         \leavevmode\phantom{aaa} Medium  & 1.34  &    [-0.15, 2.84]  &   0.078  \\         
         \leavevmode\phantom{aaa} \bf High  &\bf 3.52  & \bf   [ 1.97, 5.07] & \bf  0.000  \\            
         Work-life interference   &   &     &      \\  
         \leavevmode\phantom{aaa} Low   & Ref. & Ref.   &  Ref.  \\
         \leavevmode\phantom{aaa} Medium  & 1.15  &    [-0.32, 2.62] &   0.125  \\        
         \leavevmode\phantom{aaa} \bf High   &\bf 2.05  &\bf    [ 0.14, 3.97]  & \bf  0.036  \\            
         Job influence   & & &  \\
         \leavevmode\phantom{aaa} Low  &  1.61  &    [-0.03, 3.26]  &   0.055  \\ 
         \leavevmode\phantom{aaa} Medium  & 0.84  &    [-0.46, 2.15] &   0.204  \\
         \leavevmode\phantom{aaa} High  & Ref. &    Ref.&   Ref. \\         
         Recognition  &  & &  \\
         \leavevmode\phantom{aaa} Yes  & Ref. &    Ref.&    Ref. \\
         \leavevmode\phantom{aaa} \bf No &\bf 2.40  &  \bf  [ 0.52, 4.29] & \bf  0.012  \\       
         Job meaningfulness   & & &  \\
         \leavevmode\phantom{aaa} Low  & 1.49  &    [-0.68, 3.65]   &  0.178  \\
         \leavevmode\phantom{aaa} Medium  &  0.08  &    [-1.14, 1.30]  &   0.899  \\        
         \leavevmode\phantom{aaa} High  & Ref. &    Ref.&   Ref. \\                  
         Job insecurity  &  & &  \\
         \leavevmode\phantom{aaa} Low   & Ref. & Ref.   &  Ref.  \\
         \leavevmode\phantom{aaa} Medium  & 0.83  &    [-0.52, 2.18]   &  0.231  \\       
         \leavevmode\phantom{aaa} \bf  High  & \bf 3.71  & \bf  [ 2.22, 5.19]   &\bf  0.000  \\  
         Minority status  &  & &  \\
         \leavevmode\phantom{aaa} No  & Ref. & Ref.   &  Ref.  \\
         \leavevmode\phantom{aaa} \bf Country \& collaboration  & \bf1.93  & \bf   [ 0.39, 3.47]   & \bf 0.014  \\        
         \leavevmode\phantom{aaa} Collaboration  & -0.65  &   [-2.43, 1.13]   &  0.476  \\ 
         \leavevmode\phantom{aaa} Unknown  &  -0.74  &    [-3.61, 2.13]  &  0.612  \\         
         \hline
    \end{tabular}
    \caption{Estimated coefficients and the confidence intervals (CIs) when the level of depressive symptoms, as measured by the PHQ-9 score, is used as the dependent variable . Bold values indicates p-values $\leq$ 0.05.}
    \label{tab:PHQ9}
\end{table*}

\begin{table*}[h]
    \centering
    \begin{tabular}{|p{4,6 cm}p{2 cm}p{2.8 cm}p{2 cm}|}
        \hline
         &  \multicolumn{3}{c|}{Multivariate regression}\\
         \hline
         & Odds ratio & 95\% CI & p \\
         \hline
         Age  & & &  \\
         \leavevmode\phantom{aaa} \bf 18-29  & \bf 21.57 & \bf [2.39, 194.59]  &   \bf 0.006  \\
         \leavevmode\phantom{aaa} 30-39  & 7.83 & [0.96, 64.01]  &   0.055  \\
         \leavevmode\phantom{aaa} 40-49  &  5.73 & [0.70, 47.24]  &   0.105  \\
         \leavevmode\phantom{aaa} 50+  & Ref. & Ref. & Ref.  \\
         \leavevmode\phantom{aaa} Unknown  & 8.18 & [0.94, 71.40]  &   0.057  \\
         Gender  & & &  \\
         \leavevmode\phantom{aaa} Male  & Ref. & Ref. & Ref.  \\
         \leavevmode\phantom{aaa} Female  & 0.84 & [0.42, 1.68]   &  0.615 \\
         \leavevmode\phantom{aaa} Other  &  1.14 & [0.22, 5.78]   &  0.876  \\
         Career  & & &  \\
         \leavevmode\phantom{aaa} BA/MA student  & 1.01 & [0.13, 7.65]   &  0.994  \\
         \leavevmode\phantom{aaa} PhD  &0.79 & [0.18, 3.43] &   0.749  \\
         \leavevmode\phantom{aaa} Post-doc  &  1.42 & [0.38, 5.31]   &   0.602  \\ 
         \leavevmode\phantom{aaa} \bf Staff scientist, engineer &\bf 4.02 & \bf [1.22, 13.24]   &\bf   0.022   \\ 
         \leavevmode\phantom{aaa} Tenure track professor  &1.49 & [0.30, 7.45]  &   0.627  \\ 
         \leavevmode\phantom{aaa} Tenured professor   & Ref.  &    Ref. &    Ref.  \\         
         Working location:  onsite   & & &  \\
         \leavevmode\phantom{aaa} Yes  & Ref.  &    Ref. &    Ref.  \\
         \leavevmode\phantom{aaa}  No  &  2.22 & [0.89, 5.57] &   0.089  \\
         Uneven workload   &    &      &       \\    
         \leavevmode\phantom{aaa} Low &  Ref.  & Ref.    &  Ref.   \\
         \leavevmode\phantom{aaa} Medium & 0.65 & [0.24, 1.75] &   0.393  \\         
         \leavevmode\phantom{aaa}  High   & 1.00 & [0.38, 2.63]  &  0.995  \\            
         Work-life interference   &   &     &      \\  
         \leavevmode\phantom{aaa} Low  & Ref.  & Ref.    &  Ref.   \\
         \leavevmode\phantom{aaa} Medium  & 1.02 & [0.42, 2.46]  &   0.963  \\        
         \leavevmode\phantom{aaa} High   &1.97 & [0.74, 5.28]   &   0.176   \\            
         Job influence  & &  &  \\
         \leavevmode\phantom{aaa} \bf Low   &  \bf 2.64 &\bf [1.11, 6.26]  &   \bf 0.028  \\ 
         \leavevmode\phantom{aaa} Medium  & 1.85 & [0.95, 3.61] &   0.072  \\
         \leavevmode\phantom{aaa} High   & Ref.  &    Ref. &   Ref.  \\         
         Recognition  & &  &  \\
         \leavevmode\phantom{aaa} Yes & Ref.  &    Ref. &    Ref.  \\
         \leavevmode\phantom{aaa} \bf No  &\bf 2.90 &\bf [1.33, 6.31] & \bf  0.007  \\       
         Job meaningfulness   & & &  \\
         \leavevmode\phantom{aaa} Low  & 1.38 & [0.49, 3.88]  &  0.539  \\
         \leavevmode\phantom{aaa} Medium  &  1.09 & [0.56, 2.15]  &   0.792 \\        
         \leavevmode\phantom{aaa} High  & Ref.  &    Ref. &   Ref.  \\                  
         Job insecurity  & & &  \\
         \leavevmode\phantom{aaa} Low   & Ref.  & Ref.    &  Ref.   \\
         \leavevmode\phantom{aaa} Medium  & 1.13 & [0.43, 2.99]   & 0.808  \\       
         \leavevmode\phantom{aaa} \bf  High  & \bf 3.87 &\bf [1.62, 9.26]  &\bf 0.002  \\  
         Minority status  & & &  \\
         \leavevmode\phantom{aaa} No  & Ref.  & Ref.    &  Ref.   \\
         \leavevmode\phantom{aaa} Country \& collaboration  & 1.24 & [0.53, 2.88]   &  0.621  \\        
         \leavevmode\phantom{aaa} Collaboration   & 1.87 & [0.83, 4.20]  & 0.130   \\ 
         \leavevmode\phantom{aaa} Unknown  & 1.78 & [0.51, 6.17]   &  0.364  \\         
    \hline
    \end{tabular}
    \caption{Estimated odds ratios and confidence intervals (CIs) when an indicator variable for high desire to leave academia/collaboration (i.e. consider leaving on a weekly/daily basis) is used as the dependent variable. Bold values indicates p-values $\leq$ 0.05.}
    \label{tab:Leaving}
\end{table*}

\section{Discussion}
\label{sec:discussion}

To the best of our knowledge, the present study constitutes the first survey that investigated the prevalence of depressive symptoms as well as the desire to leave academia/the collaboration among scientists in large, international collaborations in STEM fields. 

The findings from our multivariate regression analyses indicate that several factors are associated with a higher prevalence of depressive symptoms. Specifically, early-career scientists, individuals belonging to minoritized groups, and those who experience high levels of job insecurity, uneven workload, work-life interference, and lack of recognition are more likely to exhibit higher rates of depressive symptoms compared to their more senior counterparts and those without work-related issues. Additionally, our analysis reveals that younger scientists, individuals who perceive a lack of influence and recognition in their work, and those who experience high job insecurity are more inclined to consider leaving academia or the collaboration on a regular basis. It is worth noting that our model controls for the effect of age, and thus that these results cannot be attributed solely to the fact that young individuals often report poorer mental health, particularly during and following the pandemic period.

In terms of the prevalence of depressive symptoms, the one in our sample for BA/MA and PhD students is similar to the ones reported before the pandemic by Evans et al \cite{Evans:2018}. The prevalence is slightly lower compared to the one reported for BA/MA during the peak of the pandemic by Essadek et al \cite{ESSADEK2020392}. However, the results might have been affected by the limited sample sizes for BA/MA students in our sample. 
The prevalence of depressive symptoms for a sub-population with graduate or higher levels of education in the United States before the pandemic was reported to be significantly lower compared to our sample \cite{Patel_2019}. Even when we factor in the fact that the age composition of our sample contains is relative younger than the general population, this seems insufficient to explain the observed difference in the prevalence of depressive symptoms. Two possible reasons for the higher prevalence in our sample could be the timing of the survey during the (aftermath of the) COVID-19 pandemic and/or the nature of the academic work environment. However, further research is needed to gain more insights in these aspects.

The results of the present study highlight the potential negative implications of the lack of recognition experienced by some members working in `big science' collaborations, particularly for early-career scientists. This lack of recognition can contribute to a sense of job insecurity among these researchers, as they may struggle to demonstrate their past achievements when seeking positions outside of the collaboration. Furthermore, due to the large scale of these collaborations, the number of tenure-track faculty positions is typically limited in comparison to the number of post-docs, graduate students, and other scientists involved. As demonstrated earlier, feelings of job insecurity also emerge as a significant determinant of depressive symptoms and the inclination to leave academia or the collaboration.

Furthermore, these large scientific collaborations also have unique features that can have detrimental effects on working conditions and work-life balance. The nature and scale of collaboration networks necessitate extensive coordination among multiple sites and institutions, resulting in a substantial number of meetings. Moreover, these meetings often take place outside regular business hours due to the international and multi-institutional nature of these collaborations, involving scientists from different continents. For instance, during the academic year, typically more than ten meetings are scheduled per week. Although members are not obligated to attend all scheduled meetings, the frequency and inconvenient timing of these meetings are highlighted as concerns in the open-ended responses provided in the survey.

Indeed, as reported in Fig. \ref{fig:ReasonsLeaving}, the two dominant reasons that they consider leaving academia/the collaboration are `uncertain career prospects' and `the difficulty to maintain work-life balance'. This is true for all participants including those who consider leaving academia/the collaboration only occasionally (filled bins) as well as the subgroup of participants who think about leaving more than several times a week (dashed bin-edges). 
While the act of contemplating leaving academia/the collaboration is a priori neither positive nor negative, the reasons most often indicated by participants for their inclination to leave academia/the collaboration tend to have a more negative connotation. The response `No longer interested in this type of (academic) work', which is of a more neutral or positive nature, was selected the least often compared to other reasons.

At the same time, it is also worth noting that a majority of the respondents expressed a sense of meaningfulness in their work (62.7\% of all respondents) and reported a high level of influence in their work (57.2\%). Additionally, a vast majority (more than 80 percent) of the respondents did not regularly consider leaving academia or the collaboration. These findings suggest that a significant number of scientists involved in the detection of gravitational waves perceive their work as valuable and worthwhile endeavors. Nevertheless, our study results also indicate that specific groups, particularly early-career researchers and those who lack a sense of recognition, could benefit from proactive efforts by the collaborations to address their concerns.

Regarding minoritized groups, our findings reveal that individuals who identify themselves as belonging to minoritized groups in both the collaboration and their country of residence exhibit a higher prevalence of depressive symptoms. However, we did not find statistically significant evidence indicating that scientists who identify themselves as part of minoritized groups solely within the collaboration have a higher prevalence of depressive symptoms. Similarly, there was no evidence suggesting that minoritized groups display a higher propensity to leave academia or the collaboration after controlling for confounding factors. These findings may seem somewhat counter-intuitive, considering past studies that have demonstrated the negative experiences of traditionally underrepresented minorities in STEM fields \cite{Women_physics_2019, Cech_Waidzunas_2021}. Several potential reasons could explain the lack of a strong association between minoritized status and well-being as well as the desire to change career paths. As reported in Appendix \ref{sec:Appendix_minority}, a majority of those who identify as part of minoritized groups within these collaborations responded `No' (65.1\%) when asked if they felt excluded because of their minority status. Similarly, 38.7\% of these individuals reported that their minority status did not affect their career, and 26.4\% stated that their status influenced their career both positively and negatively. Furthermore, our findings indicate that not all respondents who could identify as part of a minoritized group actually consider themselves as such. Considering the underrepresentation of women in STEM fields, including physics, it is highly likely that the majority of our female respondents can be considered part of a minoritized group, at least within the collaboration. However, only about one third of participants identifying with the `Female' and `Other' gender categories perceive themselves as part of a `Gender minority'. This suggests that some individuals within these gender categories may not necessarily feel minoritized based on their gender. Thus, within the specific collaborations examined in this study, individuals belonging to certain minoritized groups may not have had universally overwhelmingly negative experiences associated with their minority status.

This study has several limitations. First, it is important to note that this was the first well-being survey conducted by the collaborations, resulting in a relatively low response rate among collaboration members. Furthermore, participation in the survey was voluntary, which introduces the potential for self-selection bias in our sample. It is possible that individuals with work-related issues or mental health conditions were more inclined to participate, leading to an overrepresentation of such individuals in our sample. As a result, our small sample may not be fully representative of all members within the three collaborations examined in this study. Second, the lack of a comparison group limits our ability to assess how the conditions experienced by the scientists in this study compare to those of comparable groups. Ideally, we would have included scientists in academia who are not part of large international collaborations, as well as highly educated individuals outside academia, for meaningful comparisons. However, sampling these populations may present challenges as they may not have the same internal structures and connections as the large collaborations, such as LIGO, Virgo, and KAGRA. Third, all survey items, including depressive symptoms, relied on self-report measures, and thus might be subject to measurement errors. Additionally, the concept of minoritized status reflects the subjective perception of participants, which might differ from the traditional conceptualization of minority status, as discussed above. Nevertheless, we believe that understanding participants' subjective perception of belonging to a minoritized group is valuable in predicting its associations with our outcome variables that are also subjective in nature. Finally, it is important to note that this study is cross-sectional, limiting our ability to establish causal relationships between variables.

Despite these limitations, this study offers a significant and valuable contribution to the existing literature by shedding light on the challenges and opportunities associated with working in large and international collaborations. We have conducted a follow-up well-being survey in 2022, extending the invitation to scientists in eleven additional collaborations within the field of High Energy and gravitational-wave astrophysics, in addition to the three collaborations examined in this study. By monitoring the well-beings of scientists engaged in these `big science' projects on an annual basis, we aim to accumulate a more comprehensive understanding of their experiences. Furthermore, future surveys will help us assess the extent to which the results of the present study may have been influenced by the COVID-19 pandemic, as some countries were still under COVID-related restrictions during the survey period, potentially impacting the psychological well-being of certain participants. In conclusion, our findings underscore the importance of improving recognition and providing clearer job prospects, not only in reducing reported depressive symptoms but also in mitigating the likelihood of a substantial proportion of the next generation of scientists leaving academia or collaborative endeavors.

\section*{Data availability statement}

The data collected for the current study contain sensitive and potentially identifying information, and thus cannot be publicly shared. Access to the data is restricted to members of the research team, in accordance with participant consent. However, the analysis script used in this study is available upon request from the corresponding author.

\bibliography{references}

\section*{Acknowledgements}

The authors would like to express their sincere gratitude to the LIGO, Virgo, and KAGRA collaborations for their invaluable cooperation in distributing the survey and to the members of these collaborations for actively participating in the survey. The authors are deeply appreciative of their openness and willingness to contribute to this well-being study. The authors would like to thank Anke Boone for the insightful discussions. Furthermore, the authors would like to extend their gratitude to the diversity groups within the LIGO, Virgo, and KAGRA collaborations for their valuable suggestions and contributions that greatly enhanced the development of the survey. More specifically, we would like to thank (alphabetical order) Zhoujian Cao, Chian-Shu Chen, Cassidy Eassa, Raymond Frey, Chunglee Kim, Lupin C.C. Lin, Andrew Miller, Tania Regimbau, David Shoemaker, Jessica Steinlechner, Hideyuki Tagoshi, Kevin Turbang and Nami Uchikata. 

\section*{Author contributions statement}

K.J. conceived and conducted the survey, and analysed the results. M.U. provided feedback on the design of the survey. K.J. drafted main parts of the manuscript and M.U. provided critical revision for important intellectual content. All authors have read and approved the manuscript. 

\section*{Funding}

K.J. is supported by FWO-Vlaanderen via grant number 11C5720N.

\section*{Competing interests}

K.J. is a member of the Virgo collaboration, and M.U. is the spouse of a member of the LIGO collaboration. However, their involvement in these collaborations does not influence the objectivity or interpretation of the findings presented in this manuscript.

\appendix
\counterwithin{figure}{section}
\counterwithin{table}{section}
\section{Logistic regression for depressive symptoms}
\label{sec:Appendix_logitPHQ9}

BA/MA students and participants with high levels of insecurity about furture job prospects are the strongest associated with the prevalence of people with depressive symptoms (PHQ-9 $\geq$ 10). BA/MA students are 11.98 times (95\% CI: 2.37-60.53) more likely to have depressive symptoms compared to the reference group (tenured professors). In addition, PhD students are more likely to experience depressive symptoms (OR= 4.49; 95\% CI: 1.23-16.45). Feeling insecure about future job prospects increases the odds on depressive symptoms by 5.33 times (95\% CI: 2.35-12.08). Also high levels of uneven workload distribution (OR= 4.05; 95\% CI: 1.13-14.52) and lack of recognition (OR= 2.40; 95\% CI: 1.08-5.34) are linked to higher odds of depressive symptoms.

Whereas staff scientists/engineers, people not working onsite, participants with high levels of work-life interference and minoritized people (Country \& collaboration) did have significantly higher PHQ-9 scores in the multivariate linear regression in Sec. \ref{sec:results}, we did not find increased odds at having depressive symptoms (PHQ-9 $\geq$ 10) among these subgroups when a dichotomous dependent variable was used.

\begin{table*}[h]
    \centering
    \begin{tabular}{|p{4,6 cm}p{2 cm}p{2.75 cm}p{2 cm}|}
        \hline
         & \multicolumn{3}{c|}{Multivariate regression}\\
         \hline
         & Odds ratio & 95\% CI & p \\
         \hline
         Age &  & &  \\
         \leavevmode\phantom{aaa} 18-29  &  0.78 & [0.17, 3.65]  & 0.754    \\
         \leavevmode\phantom{aaa} 30-39  &  0.60 & [0.16, 2.20]  & 0.441   \\
         \leavevmode\phantom{aaa} 40-49  & 0.72 & [0.19, 2.75]  & 0.627  \\
         \leavevmode\phantom{aaa} 50+  & Ref. & Ref. & Ref.  \\
         \leavevmode\phantom{aaa} Unknown  & 0.61 & [0.15, 2.54]  & 0.501 \\
         Gender  & & &  \\
         \leavevmode\phantom{aaa} Male  & Ref. & Ref. & Ref.  \\
         \leavevmode\phantom{aaa} Female  &  1.33 & [0.69, 2.56]  & 0.401 \\
         \leavevmode\phantom{aaa} Other  & 2.12 & [0.54, 8.37]  & 0.285\\
         Career  & & &  \\
         \leavevmode\phantom{aaa} \bf BA/MA student  &\bf 11.98 &\bf  [2.37, 60.53]    & \bf 0.003  \\
         \leavevmode\phantom{aaa} \bf PhD  &\bf 4.49 &\bf  [1.23, 16.45]    & \bf 0.023   \\
         \leavevmode\phantom{aaa} Post-doc  &   1.39 & [0.45, 4.34]    & 0.569 \\ 
         \leavevmode\phantom{aaa}  Staff scientist, engineer   &   2.28 & [0.73, 7.09]    & 0.155 \\ 
         \leavevmode\phantom{aaa} Tenure track professor &  1.38 & [0.28, 6.85]    & 0.694  \\ 
         \leavevmode\phantom{aaa} Tenured professor   & Ref.  &    Ref. &    Ref.  \\         
         Working location: onsite   & & &  \\
         \leavevmode\phantom{aaa} Yes  & Ref.  &    Ref. &    Ref.  \\
         \leavevmode\phantom{aaa}  No  &  1.74 & [0.71, 4.22]    & 0.224  \\
         Uneven workload   &    &      &       \\    
         \leavevmode\phantom{aaa} Low    & Ref.  & Ref.    &  Ref.   \\
         \leavevmode\phantom{aaa} Medium &  2.01 & [0.62, 6.53]  & 0.246   \\         
         \leavevmode\phantom{aaa} \bf High   & \bf4.05 &\bf [1.13, 14.52]  &\bf 0.032 \\            
         Work-life interference   &   &     &      \\  
         \leavevmode\phantom{aaa} Low  & Ref.  & Ref.    &  Ref.   \\
         \leavevmode\phantom{aaa} Medium  & 1.33 & [0.51, 3.44]  & 0.557 \\        
         \leavevmode\phantom{aaa} High  &  2.37 & [0.78, 7.21]  & 0.130 \\            
         Job influence   & & &  \\
         \leavevmode\phantom{aaa}  Low   &   2.28 & [1.00, 5.22]  & 0.051  \\ 
         \leavevmode\phantom{aaa} Medium  & 1.56 & [0.78, 3.10]  & 0.206   \\
         \leavevmode\phantom{aaa} High   & Ref.  &    Ref. &   Ref.  \\         
         Recognition   & & &  \\
         \leavevmode\phantom{aaa} Yes  & Ref.  &    Ref. &    Ref.  \\
         \leavevmode\phantom{aaa} \bf No  &\bf 2.40 &\bf [1.08, 5.34]  &\bf 0.032  \\       
         Job meaningfulness   & & &  \\
         \leavevmode\phantom{aaa} Low  & 0.77 & [0.23, 2.50]  & 0.658 \\
         \leavevmode\phantom{aaa} Medium  &   0.87 & [0.45, 1.67]  & 0.672  \\        
         \leavevmode\phantom{aaa} High  & Ref.  &    Ref. &   Ref.  \\                  
         Job insecurity   & & &  \\
         \leavevmode\phantom{aaa} Low   & Ref.  & Ref.    &  Ref.   \\
         \leavevmode\phantom{aaa} Medium  &  1.33 & [0.52, 3.43]  & 0.550   \\       
         \leavevmode\phantom{aaa} \bf  High  & \bf  5.33 &\bf [2.35, 12.08]  &\bf 0.000 \\  
         Minority status   & & &  \\
         \leavevmode\phantom{aaa} No  & Ref.  & Ref.    &  Ref.   \\
         \leavevmode\phantom{aaa} Country \& collaboration  &  1.01 & [0.50, 2.02]  & 0.977 \\        
         \leavevmode\phantom{aaa} Collaboration  & 0.56 & [0.22, 1.46]  & 0.238  \\ 
         \leavevmode\phantom{aaa} Unknown &  0.31 & [0.06, 1.69]  & 0.175 \\         
    \hline
    \end{tabular}
    \caption{Estimated odds ratios and confidence intervals (CIs) when an indicator variable for the presence of depressive symptoms (PHQ-9 $\geq$ 10) is used as the dependent variable. Bold values indicates p-values $\leq$ 0.05.}
    \label{tab:Depressed-logit}
\end{table*}

\section{Minority status}
\label{sec:Appendix_minority}

\begin{figure}
    \centering
    \includegraphics[width=\linewidth]{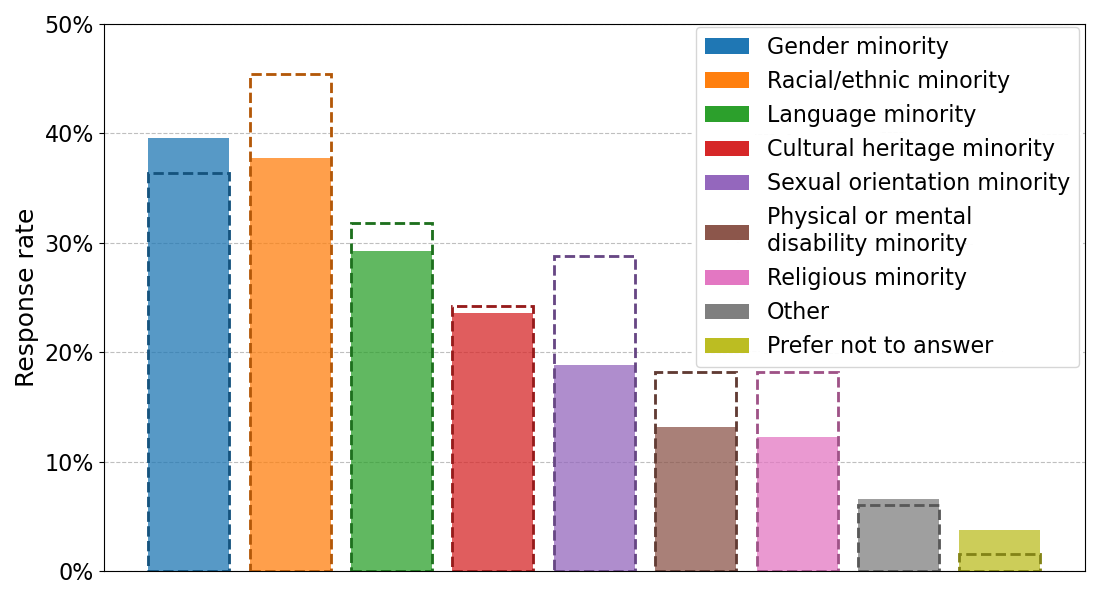}
    \caption{Histogram indicating the different minoritized groups people consider to be part of. On average people belonging to minoritized groups belong to 1.8 minoritized groups. The shown percentages are with respect to the number of people who responded to this question.
    The dashed bin edges represent the response rates of the 66 participants who consider themselves to be part of a minoritized groups both in their country of residence as well as the collaboration. These participants indicate to belong on average to 2.1 minoritized groups.}
    \label{fig:MinorityCategories}
\end{figure}

To gain further insight in the status of minoritized participants, they were asked to which minoritized group they were belonging, which is represented in Fig. \ref{fig:MinorityCategories}.   
In total 196 responses were collected for 106 participants whom indicated to belong to a minoritized group.
The answers from the 66 participants who belong to minoritized groups both in their country of residence as well as the collaboration or indicated by the dashed bin edges in Fig. \ref{fig:MinorityCategories}.
Furthermore they were asked whether they felt that being part of a minoritized group altered their career opportunities to which they could answer (1) Yes, (mainly) negatively (29.2\% of minoritized participants),(2) Yes, (mainly) positively (1.9\%),(3) Yes, sometimes negatively, sometimes positively (26.4\%), (4) No  (38.7\%) and (5) Prefer not to answer (3.8\%). Finally they were asked whether they felt excluded within the collaboration where people could indicate multiple answers from the following options (1) Yes, work related (21.7\% of minoritized participants), (2) Yes, at social events at collaboration meetings or with colleagues (17.0\%), (3) No (65.1\%) (4) Prefer not to answer (4.7\%). Note that the total percentage is larger than 100\% since 9 participants responded with both option (1) as well as (2).

\section{Recognition}
\label{sec:Appendix_recognition}

In Tab. \ref{tab:PrevalenceRecognition} we present the uncorrected fractions of participants who reported a lack of recognition both inside as well as outside the collaboration by sample characteristics. The fraction across all respondents is 13.10\% (95\% CI: 9.94\%-16.82\%).

\begin{table*}[h]
    \centering
    \begin{tabular}{|p{4,5 cm}p{5 cm}|}
        \hline
         &  \multicolumn{1}{c|}{Prevalence}\\
         \hline
         & \makecell[l]{Lack of recognition \\ (inside and outside collaboration)}\\ \hline
          Total & 13.10\%   \\
          Age &    \\
         \leavevmode\phantom{aaa} 18-29 & 14.91\%  \\
         \leavevmode\phantom{aaa} 30-39 & 15.60\%     \\
         \leavevmode\phantom{aaa} 40-49 & 11.86\% \\
         \leavevmode\phantom{aaa} 50+ &  5.45\% \\
         \leavevmode\phantom{aaa} Unknown & 13.33\%   \\
         Gender &   \\
         \leavevmode\phantom{aaa} Male &   13.19\%   \\
         \leavevmode\phantom{aaa} Female &  13.86\%    \\
         \leavevmode\phantom{aaa} Other &   8.70\%   \\
         Career &  \\
         \leavevmode\phantom{aaa} BA/MA student &  9.52\%   \\
         \leavevmode\phantom{aaa} PhD &  15.62\%   \\
         \leavevmode\phantom{aaa} Post-doc & 20.69\%  \\
         \leavevmode\phantom{aaa} Staff scientist, engineer & 15.38\%   \\
         \leavevmode\phantom{aaa} Tenure track professor &  10.71\%   \\
         \leavevmode\phantom{aaa} Tenured professor & 4.00\%   \\ 
         Working location: onsite  &  \\
         \leavevmode\phantom{aaa} Yes & 12.31\%  \\
         \leavevmode\phantom{aaa}  No &  13.25\%  \\
         Uneven workload  &             \\    
         \leavevmode\phantom{aaa} Low &  5.00\%      \\
         \leavevmode\phantom{aaa} Medium & 12.03\%    \\         
         \leavevmode\phantom{aaa}  High & 16.18\%   \\            
         Work-life interference  &          \\  
         \leavevmode\phantom{aaa} Low & 6.10\%     \\
         \leavevmode\phantom{aaa} Medium & 10.10\%   \\        
         \leavevmode\phantom{aaa}  High & 24.30\%    \\               
         Job influence   &  \\
         \leavevmode\phantom{aaa} Low & 16.98\%      \\
         \leavevmode\phantom{aaa} Medium & 19.66\%  \\
         \leavevmode\phantom{aaa} High & 8.81\%   \\         
         Recognition   &    \\
         \leavevmode\phantom{aaa} Yes & -  \\         
         \leavevmode\phantom{aaa}  No &  -  \\           
         Job meaningfulness  &  \\
         \leavevmode\phantom{aaa} Low & 45.71\%   \\
         \leavevmode\phantom{aaa} Medium & 19.47\%  \\         
         \leavevmode\phantom{aaa} High & 5.62\%  \\            
         Job insecurity  &   \\
         \leavevmode\phantom{aaa} Low &  5.45\%       \\
         \leavevmode\phantom{aaa} Medium &  8.24\%  \\         
         \leavevmode\phantom{aaa}  High &  24.49\% \\           
         Minority status &  \\
         \leavevmode\phantom{aaa} No & 9.96\%  \\
         \leavevmode\phantom{aaa} Country \& collaboration & 22.73\%  \\        
         \leavevmode\phantom{aaa} Collaboration & 15.00\%   \\ 
         \leavevmode\phantom{aaa} Unknown & 20.00\%    \\         
\hline
    \end{tabular}
    \caption{Fraction of respondents reporting a lack of recognition both inside as well as outside the collaboration.}
    \label{tab:PrevalenceRecognition}
\end{table*}

\end{document}